%% file: sample-sigconf.tex
\documentclass[sigconf]{acmart}
\usepackage{booktabs} % For formal tables

\setcopyright{none}
\acmDOI{-}

\acmISBN{-}

\acmConference[JCDL2017]{ACM/IEEE Joint Conference on Digital Libraries}{June 2017}{Toronto CA} 
\acmYear{2017}
\copyrightyear{2017}

\acmPrice{-}

\begin{document}
\title{Exploring Features for Predicting Policy Citations}
%\subtitle{Extended Abstract}

% use '\and' if you need 'another row' of author name
\author{Christian Bailey}
\affiliation{%
	\institution{Northern Illinois University}
}
\email{cbailey11@niu.edu}

\author{Bharat Kale}
\affiliation{%
	\institution{Northern Illinois University}
}
\email{bkale@niu.edu}

\author{Jamieson Walker}
\affiliation{%
	\institution{Northern Illinois University}
}
\email{jwalker21@niu.edu}

\author{Harish Varma Siravuri}
\affiliation{%
	\institution{Northern Illinois University}
}
\email{hsiravuri@niu.edu}
       
\author{Hamed Alhoori}
\affiliation{%
	\institution{Northern Illinois University}
}
\email{alhoori@niu.edu}

\author{Michael E. Papka}
\affiliation{%
	\institution{Argonne National Laboratory and Northern Illinois University}
}
\email{papka@niu.edu}

\renewcommand{\shortauthors}{C. Bailey et al.}

\begin{abstract}
In this study we performed an initial investigation and evaluation of altmetrics and their relationship with public policy citation of research papers. We examined methods for using altmetrics and other data to predict whether a research paper is cited in public policy and applied receiver operating characteristic curve on various feature groups in order to evaluate their potential usefulness. From the methods we tested, classifying based on tweet count provided the best results, achieving an area under the ROC curve of 0.91.
\end{abstract}

% We no longer use \terms command


%% Used in some conference proceedings e.g. sigplan and sigchi
%   \caption{This is a teaser}

\maketitle

\input{samplebody-conf}

\bibliographystyle{ACM-Reference-Format}
\bibliography{sigproc} 

\end{document}

%% file: samplebody-conf.tex
\section{BACKGROUND}
\subsection{Introduction}
The growth of social media in the academic community has enabled scholars to develop new methods to evaluate the impact of research. Historically, evaluations of the impact of research have been limited to the reception by the scholarly community. However, with the advent of altmetrics we are able to track the social impact of research \cite{priem2012altmetrics}\cite{Alhoori2014} . For example, Ding et al. \cite{ding2009perspectives} explored the use of social media tagging as it relates to scholarly works.

Policy documents have a vital role in generating demand for scientific innovation \cite{Edler2007}. Haunschild and Bornmann \cite{haunschild2016many} study the relation between Web of Science fields and the researches' use in public policy and found that less than 2\% of every category is cited in public policy. Orduna-Malea, Thelwall, and Kousha \cite{Orduna-Malea2017} explored the relationship between citations in patents and technological impact and found that the number of patents citing a resource indicates the technological capacity or relevance of that resource. Winterfeldt \cite{VonWinterfeldt2013} presented a framework to bridge the gap between science and decision making in the policy sphere. 

To better understand the possibilities of altmetrics, we conducted an exploratory study to determine the potential of social media data for creating valuable models to describe the use of research articles in public policy. 

\subsection{Collection}
The primary source of data for our analysis was a database dump from altmetric.com \cite{adie2013altmetric}. The dataset, which is from June 4th 2016, consists of 5.2 million articles. We separated articles into 2 classes: papers cited in public policy documents and papers not cited in such documents. Policy documents, as used in this paper and by altmetric.com, currently includes mostly policy published by medical organizations. The dataset comprised 89,350 research articles referenced in policy documents and 5,097,207 articles not referenced in that context. 

We drew on the altmetric.com dataset for meta-info about each research paper, specifically journal, publisher, and Scopus subject information, as well as social media activity. To augment our dataset, we collected the citation counts for articles of interest. Initially, we collected citation counts for all the policy documents and 120,000 of the non-policy documents from the Thomson Reuters Web of Science. In addition, we collected journal impact factors for 9,000 journals.
% * <alhoori@hotmail.com> 2017-04-26T11:08:13.426Z:
% 
% > domain
% This is not very clear
% 
% ^.
% * <alhoori@hotmail.com> 2017-02-12T23:12:44.238Z:
% 
% > In our work the altmetric.com dataset was used for meta-info and social media activity for research papers.
% Not very clear 
% 
% ^ <z1787641@students.niu.edu> 2017-02-12T23:42:15.559Z.
\subsection{Feature Selection and Filtering}
We selected four groups of features to evaluate: meta-info consisting of journal, publisher, and Scopus subject; social media information consisting of unique user mentions on Facebook, Google Plus, Twitter, Reddit, and Stackoverflow; traditional citation counts; and mention counts from several online platforms. We filtered the social media mentions to consider only those that had occurred before the policy citation of any given article. We treated the count features as simple numeric variables, whereas for meta-info and unique users we used a collection of boolean variables for each value.
% * <alhoori@hotmail.com> 2017-04-26T11:13:31.542Z:
% 
% > We stored the two former feature sets each represented as collections of boolean features, and the remainder each as a count feature.
% Not clear 
% 
% ^.
\section{Methods}
We used two methods to explore the efficacy of our selected feature sets for predicting policy citations. The first approach we used was to perform linear regression on our various feature sets, and the second involved training classifiers and evaluating various classification metrics. In both instances, we used the Scikit-Learn \cite{scikit-learn} software package to develop our models.

% * <christianbailey0556@gmail.com> 2017-02-12T23:10:16.739Z:
% 
% we can comment like this
% 
% 
% ^ <alhoori@hotmail.com> 2017-02-12T23:14:49.486Z.

\subsection{Linear Regression and Correlation}
To analyze the relations between our feature sets and policy citations, we used least squares regression and calculated the coefficients of determination ($r^2$) for several relationships. We related each of citation count, unique users, meta-info, and mention counts on a number of platforms including Twitter, Google Plus, Facebook, Wikipedia, Mendeley, blogs, and Stackoverflow to two targets: policy citation presence, i.e., a boolean target, and policy citation count, which is the number of policy documents citing an article.
% * <alhoori@hotmail.com> 2017-02-13T00:51:04.648Z:
% 
% > unique users,
% in a paper or policy doc or something else?
% 
% ^.

We list the coefficient of determination for each relation in the Table 1. While overall, the correlations are relatively weak, we observe a wide range in values, so it is clear that some features have a stronger bearing on whether a document is cited in policy documents.
\begin{table}[!h]
\centering
\caption{$r^2$ for features regressed on Presence and Count}
\begin{tabular}{|r|c|c|} \hline
&Policy Presence&Policy Count\\ \hline
Citation Count & $0.01118$ & $0.01557$\\ \hline
Unique Users & $0.32163$ & $0.15005$\\ \hline
Meta-info & $0.41748$ & $0.27011$\\ \hline
Blog Posts & $0.04242$ & $0.01449$ \\ \hline
Mendeley Readers & $0.03977$ & $0.03830$ \\ \hline
Wikipedia & $0.01152$ & $0.00617$\\ \hline
Tweet Count & $0.01460$ & $0.00612$ \\ \hline
Facebook Posts & $0.00771$ & $0.00361$ \\ \hline
Google Plus & $0.00265$ & $0.00012$ \\ \hline
Stackoverflow & $0.00150$ & $0.00036$ \\
\hline\end{tabular}
\end{table}
\subsection{Classification}
\begin{table}[!h]
\centering
\caption{Area under ROC curve with a 90\% CI}
\begin{tabular}{|r|c|} \hline
&ROC AUC using Bernoulli NB\\ \hline
Citation Count & $0.57\pm 0.0$\\ \hline
Unique Users & $0.71\pm 0.05$\\ \hline
Meta-info & $0.81\pm 0.04$\\ \hline
Tweet Count & $0.91\pm 0.01$\\ \hline
Facebook Posts & $0.62\pm 0.04$\\ \hline
Blog Posts & $0.54\pm 0.06$\\ \hline
Mendeley Readers & $0.54\pm 0.01$\\ \hline
Wikipedia & $0.52\pm 0.01$\\ \hline
Google Plus & $0.52\pm 0.01$\\ \hline
Stackoverflow & $0.50\pm 0.01$\\
\hline\end{tabular}
\end{table}
\sloppy To determine which, if any, feature sets provide the best predictions for citations in policy, we evaluated Bernoulli Naive Bayes using unique user mentions, meta-info, citation count, and mention counts on platforms including several social media sites, Wikipedia, Mendeley, and blogs. Instead of the standard method of counting policy citations, we used binary labels for the purpose of classification: either cited in policy or not cited in policy.
 
We performed a ten-fold cross validation evaluation on each of the stated feature sets. From this evaluation, we determined the area under the receiver operating characteristic curve when classifying the presence of a policy citation with each feature set, as shown in Table 2. We found that citation and mention counts performed poorly. However, classifiers using unique users and journal meta-info performed better.

\section{Conclusions and Future Work}

In this initial study, we took a handful of alternative metrics and classic metrics for research papers to examine how they relate to the use of scholarly research in policy documents. We found that citations to be very poor at predicting research use in public policy. However, altmetrics grouped in specific ways, such as unique users and meta-info, show better potential.

With this study, we have discovered some promising directions for additional research and discounted a handful of other avenues of study. Moving forward, we plan to develop a more in-depth study of correlations between altmetrics and the use of research in public policy. We plan to focus on applying clustering algorithms and methods to altmetrics to determine which clusters produce the most accurate predictions. We will continue to compare our results from working with altmetrics to similar tests based on more classic ways to evaluate research articles and determine the value of given scholarly research papers. We plan to develop a model that will provide accurate and timely predictions pertaining to whether any given scholarly research will be credited in public policy documents.
 
\section{Acknowledgements}
MEP was supported in part by the Office of Advanced Scientific Computing Research, Office of Science, U.S. Department of Energy, under Contract DE-AC02-06CH11357.

%% file: sample-sigconf.bbl

\begin{thebibliography}{00}

%%% the Web address.

\ifx \showCODEN    \undefined \def \showCODEN     #1{\unskip}     \fi
\ifx \showDOI      \undefined \def \showDOI       #1{#1}\fi
\ifx \showISBNx    \undefined \def \showISBNx     #1{\unskip}     \fi
\ifx \showISBNxiii \undefined \def \showISBNxiii  #1{\unskip}     \fi
\ifx \showISSN     \undefined \def \showISSN      #1{\unskip}     \fi
\ifx \showLCCN     \undefined \def \showLCCN      #1{\unskip}     \fi
\ifx \shownote     \undefined \def \shownote      #1{#1}          \fi
\ifx \showarticletitle \undefined \def \showarticletitle #1{#1}   \fi
\ifx \showURL      \undefined \def \showURL       {\relax}        \fi
% invisible to TeX
\providecommand\bibfield[2]{#2}
\providecommand\bibinfo[2]{#2}
\providecommand\natexlab[1]{#1}
\providecommand\showeprint[2][]{arXiv:#2}

\bibitem[\protect\citeauthoryear{Adie and Roe}{Adie and Roe}{2013}]%
        {adie2013altmetric}
\bibfield{author}{\bibinfo{person}{Euan Adie} {and} \bibinfo{person}{William
  Roe}.} \bibinfo{year}{2013}\natexlab{}.
\newblock \showarticletitle{Altmetric: enriching scholarly content with
  article-level discussion and metrics}.
\newblock \bibinfo{journal}{{\em Learned Publishing\/}} \bibinfo{volume}{26},
  \bibinfo{number}{1} (\bibinfo{year}{2013}), \bibinfo{pages}{11--17}.
\newblock


\bibitem[\protect\citeauthoryear{Alhoori and Furuta}{Alhoori and
  Furuta}{2014}]%
        {Alhoori2014}
\bibfield{author}{\bibinfo{person}{Hamed Alhoori} {and}
  \bibinfo{person}{Richard Furuta}.} \bibinfo{year}{2014}\natexlab{}.
\newblock \showarticletitle{Do Altmetrics Follow the Crowd or Does the Crowd
  Follow Altmetrics?}. In \bibinfo{booktitle}{{\em Proceedings of the 14th
  ACM/IEEE-CS Joint Conference on Digital Libraries}} {\em
  (\bibinfo{series}{JCDL '14})}. \bibinfo{publisher}{IEEE Press},
  \bibinfo{address}{Piscataway, NJ, USA}, \bibinfo{pages}{375--378}.
\newblock
\showISBNx{978-1-4799-5569-5}
\showURL{%
\url{http://dl.acm.org/citation.cfm?id=2740769.2740833}}


\bibitem[\protect\citeauthoryear{Ding, Jacob, Zhang, Foo, Yan, George, and
  Guo}{Ding et~al\mbox{.}}{2009}]%
        {ding2009perspectives}
\bibfield{author}{\bibinfo{person}{Ying Ding}, \bibinfo{person}{Elin~K. Jacob},
  \bibinfo{person}{Zhixiong Zhang}, \bibinfo{person}{Schubert Foo},
  \bibinfo{person}{Erjia Yan}, \bibinfo{person}{Nicolas~L. George}, {and}
  \bibinfo{person}{Lijiang Guo}.} \bibinfo{year}{2009}\natexlab{}.
\newblock \showarticletitle{Perspectives on social tagging}.
\newblock \bibinfo{journal}{{\em Journal of the American Society for
  Information Science and Technology\/}} \bibinfo{volume}{60},
  \bibinfo{number}{12} (\bibinfo{year}{2009}), \bibinfo{pages}{2388--2401}.
\newblock
\showISSN{1532-2890}
\showDOI{%
\url{https://doi.org/10.1002/asi.21190}}


\bibitem[\protect\citeauthoryear{Edler and Georghiou}{Edler and
  Georghiou}{2007}]%
        {Edler2007}
\bibfield{author}{\bibinfo{person}{Jakob Edler} {and} \bibinfo{person}{Luke
  Georghiou}.} \bibinfo{year}{2007}\natexlab{}.
\newblock \showarticletitle{{Public procurement and innovation—Resurrecting
  the demand side}}.
\newblock \bibinfo{journal}{{\em Research Policy\/}} \bibinfo{volume}{36},
  \bibinfo{number}{7} (\bibinfo{date}{sep} \bibinfo{year}{2007}),
  \bibinfo{pages}{949--963}.
\newblock
\showISBNx{0048-7333}
\showISSN{00487333}
\showDOI{%
\url{https://doi.org/10.1016/j.respol.2007.03.003}}
\showeprint[arxiv]{arXiv:1011.1669v3}


\bibitem[\protect\citeauthoryear{Haunschild and Bornmann}{Haunschild and
  Bornmann}{2016}]%
        {haunschild2016many}
\bibfield{author}{\bibinfo{person}{Robin Haunschild} {and}
  \bibinfo{person}{Lutz Bornmann}.} \bibinfo{year}{2016}\natexlab{}.
\newblock \showarticletitle{How many scientific papers are mentioned in
  policy-related documents? An empirical investigation using Web of Science and
  Altmetric data}.
\newblock \bibinfo{journal}{{\em Scientometrics\/}} (\bibinfo{year}{2016}),
  \bibinfo{pages}{1--8}.
\newblock


\bibitem[\protect\citeauthoryear{Orduna-Malea, Thelwall, and
  Kousha}{Orduna-Malea et~al\mbox{.}}{2017}]%
        {Orduna-Malea2017}
\bibfield{author}{\bibinfo{person}{Enrique Orduna-Malea}, \bibinfo{person}{Mike
  Thelwall}, {and} \bibinfo{person}{Kayvan Kousha}.}
  \bibinfo{year}{2017}\natexlab{}.
\newblock \showarticletitle{{Web citations in patents: Evidence of
  technological impact?}}
\newblock \bibinfo{journal}{{\em Journal of the Association for Information
  Science and Technology\/}} (\bibinfo{year}{2017}).
\newblock
\showDOI{%
\url{https://doi.org/10.1002/asi.23821}}


\bibitem[\protect\citeauthoryear{Pedregosa, Varoquaux, Gramfort,
  et~al\mbox{.}}{Pedregosa et~al\mbox{.}}{2011}]%
        {scikit-learn}
\bibfield{author}{\bibinfo{person}{F. Pedregosa}, \bibinfo{person}{G.
  Varoquaux}, \bibinfo{person}{A. Gramfort}, {and} \bibinfo{person}{others}.}
  \bibinfo{year}{2011}\natexlab{}.
\newblock \showarticletitle{Scikit-learn: Machine Learning in {P}ython}.
\newblock \bibinfo{journal}{{\em J. Mach. Learn. Res.\/}}  \bibinfo{volume}{12}
  (\bibinfo{year}{2011}), \bibinfo{pages}{2825--2830}.
\newblock


\bibitem[\protect\citeauthoryear{Priem, Piwowar, and Hemminger}{Priem
  et~al\mbox{.}}{2012}]%
        {priem2012altmetrics}
\bibfield{author}{\bibinfo{person}{Jason Priem}, \bibinfo{person}{Heather~A.
  Piwowar}, {and} \bibinfo{person}{Bradley~M. Hemminger}.}
  \bibinfo{year}{2012}\natexlab{}.
\newblock \showarticletitle{Altmetrics in the wild: Using social media to
  explore scholarly impact}.
\newblock \bibinfo{journal}{{\em CoRR\/}}  \bibinfo{volume}{abs/1203.4745}
  (\bibinfo{year}{2012}).
\newblock
\showURL{%
\url{http://arxiv.org/abs/1203.4745}}


\bibitem[\protect\citeauthoryear{von Winterfeldt}{von Winterfeldt}{2013}]%
        {VonWinterfeldt2013}
\bibfield{author}{\bibinfo{person}{Detlof von Winterfeldt}.}
  \bibinfo{year}{2013}\natexlab{}.
\newblock \showarticletitle{{Bridging the gap between science and decision
  making}}.
\newblock \bibinfo{journal}{{\em Proceedings of the National Academy of
  Sciences\/}} \bibinfo{volume}{110}, \bibinfo{number}{Supplement{\_}3}
  (\bibinfo{date}{aug} \bibinfo{year}{2013}), \bibinfo{pages}{14055--14061}.
\newblock
\showISBNx{1091-6490}
\showISSN{0027-8424}
\showDOI{%
\url{https://doi.org/10.1073/pnas.1213532110}}


\end{thebibliography}
